\documentclass[11pt]{article}
\usepackage{setspace}
\usepackage[margin=1in]{geometry}
\usepackage[T1]{fontenc}
\usepackage[utf8]{inputenc}
\usepackage{amsmath,amssymb,amsthm}
\usepackage{booktabs}
\usepackage{graphicx}
\usepackage{float}
\usepackage{algorithm}
\usepackage{algpseudocode}
\usepackage[authoryear,round]{natbib}
\usepackage[colorlinks=true,linkcolor=blue,citecolor=blue,urlcolor=blue]{hyperref}

\newtheorem{proposition}{Proposition}
\newtheorem{theorem}{Theorem}

\newcommand{\E}{\mathbb E}
\newcommand{\Prb}{\mathbb P}
\newcommand{\ind}{\mathbf 1}
\newcommand{\expit}{\operatorname{expit}}

\title{Direct Learning of Treatment-Benefit Rankings for Restricted Mean Survival Time in Randomized Trials with Censoring}

\author{Lingli Yang\thanks{Corresponding author: \texttt{Lingli.Yang@takeda.com}}, Tian Chen, Han Zhu, Xuzhi Wang, Bingxia Wang\\[2pt]
\small Statistical and Quantitative Science Department, Takeda Pharmaceutical Company, Ltd.,\\
\small Cambridge, Massachusetts, USA}

\date{}
\begin{document}

\maketitle
\begin{abstract}
In precision medicine, treatment decisions often depend more on identifying which patients are most likely to benefit from treatment than on accurately estimating benefit for every patient. However, existing methods estimate patient-specific treatment effects and derive patient rankings as a secondary step.
We propose a direct ranking approach for conditional RMST treatment benefit in randomized trials with right-censored outcomes. We first construct a censoring-adjusted orthogonal RMST pseudo-outcome whose conditional expectation equals the true conditional RMST difference. Then, we compare patients pairwise and optimize a smooth ranking loss that directly targets treatment-benefit ordering. The pairwise criterion incorporates orthogonal corrections for estimation of the event and censoring distributions.
We show that the population minimizer of the proposed loss induces the same ordering as the true conditional RMST treatment benefit and that the resulting estimating equation is Neyman-orthogonal to nuisance survival and censoring models. In simulation studies with nonlinear treatment-effect heterogeneity and moderate to heavy censoring, the proposed approach improved rank correlation and treatment-benefit enrichment relative to both plug-in RMST estimation and regression of the same RMST pseudo-outcome. Application to a randomized breast cancer trial demonstrated improved identification of patients with greater treatment benefit. In conclusion, the proposed framework combines orthogonal survival learning with pairwise ranking and may be particularly useful in precision medicine settings where treatment prioritization and patient selection are the primary goals.
\medskip

\end{abstract}

\section{Introduction}

Treatment effects may vary across patients in a randomized clinical trial. For a survival endpoint, treatment benefit can be measured by the difference in restricted mean survival time (RMST). The RMST through a prespecified time \(t^\star\) is the expected amount of time that a patient remains alive and event free during the interval from 0 to \(t^\star\). For patient \(i\), let \(X_i\) denote the baseline characteristics measured before treatment. Under treatment \(a\), define
\[
\operatorname{RMST}_a(t^\star\mid X_i)
=
\int_0^{t^\star} S_a(u\mid X_i)\,du,
\]
where \(a=1\) denotes the experimental treatment, \(a=0\) denotes the control treatment, and \(S_a(u\mid X_i)\) is the probability that a patient with baseline characteristics \(X_i\) remains event free beyond time \(u\) under treatment \(a\). We define the conditional RMST treatment benefit as
\[
\Delta(X_i)
=
\operatorname{RMST}_1(t^\star\mid X_i)
-
\operatorname{RMST}_0(t^\star\mid X_i).
\]
Thus, \(\Delta(X_i)\) is the expected gain in event-free survival time through \(t^\star\) under the experimental treatment for a patient with baseline characteristics \(X_i\). It has units of time, and a positive value favors the experimental treatment. In some applications, the ordering of these benefits is of greater interest than their exact numerical values. For example, estimates of 5.3 and 4.0 months for two patients whose true benefits are 6 and 2 months are inaccurate in magnitude but give the correct order. We study an estimator that targets this ordering directly.

Ranks and pairwise comparisons have a long history in clinical trials. The Wilcoxon rank sum statistic aggregates comparisons between patients in two randomized groups. Extensions include global rank procedures for multiple endpoints \citep{obrien1984,li2023multivariate}, pairwise evaluation of mortality and longitudinal outcomes \citep{finkelstein1999}, generalized pairwise comparisons \citep{buyse2010}, and the win ratio for prioritized composite endpoints \citep{pocock2012}. \citet{li2023multivariate} considered joint ranking of all endpoints. Pairwise methods have also been developed for censored time-to-event outcomes \citep{pan2026}. These methods compare patient outcomes under different estimands. Here, pairwise comparisons are used to order conditional treatment benefits.

We use pairwise comparisons to rank treatment benefit. For patients $i$ and $j$, the target comparison is
\[
\Delta(X_i)
\quad\text{and}\quad
\Delta(X_j).
\]
Prognosis alone cannot determine this comparison: a patient with longer expected survival under both treatments need not have the larger treatment benefit. Benefit is also unobserved because each patient receives only one treatment, and the event time may be censored. Methods for heterogeneous treatment effects address these problems by estimating conditional treatment effects from randomized trial data. SurvITE models treatment-specific hazards using representation learning \citep{curth2021}, causal survival forests use forest-based procedures \citep{cui2023}, and orthogonal survival learners estimate several survival contrasts, including RMST differences, while accounting for censoring \citep{frauen2025}. These methods can produce a ranking by sorting the estimated treatment effects. However, our goal is to model the relative orders directly.

Recent work has treated the ordering itself as the estimation target. PairNet uses observed patient pairs to estimate individual treatment effects \citep{nagalapatti2024}, and Rank Learner uses an orthogonal pairwise loss to learn treatment effect rankings \citep{arno2026}. Rank-weighted average treatment effects and targeting operator characteristic curves provide tools for evaluating whether higher scores correspond to larger treatment effects \citep{yadlowsky2025}. We adapt orthogonal treatment-effect ranking to randomized trials with right-censored survival outcomes and use conditional RMST difference as the treatment-effect measure.

For each patient, we construct an RMST pseudo outcome that adjusts for treatment assignment and censoring and has conditional mean \(\Delta(X_i)\). Let \(r(x)\) denote a score calculated from baseline characteristics, with larger values indicating larger expected RMST benefit. The score is used only for ordering and has no interpretation in units of survival time. For patients \(i\) and \(j\), the pairwise criterion favors
\[
r(X_i)>r(X_j)
\]
when the adjusted survival information favors \(\Delta(X_i)>\Delta(X_j)\). The score is estimated by combining these comparisons across patient pairs.

The pairwise criterion includes an orthogonal correction for estimation of the event and censoring distributions. We show that its population minimizer orders patients by \(\Delta(x)\) and that its estimating equation is Neyman-orthogonal. The simulations compare direct ranking with a plug-in RMST estimator and with regression of the same RMST pseudo outcome. Performance is measured by rank agreement and by the treatment benefit among patients with the highest scores. An application to the German Breast Cancer Study Group 2 trial illustrates the analysis with recurrence-free survival data.

\section{Method}

Consider a randomized trial with baseline covariates, treatment assignment, follow-up time, and an event indicator. For a prespecified RMST restriction time, the analysis assigns each baseline profile a score, with larger scores indicating larger predicted treatment benefit.

\subsection{Treatment effect}

Let $X$ denote the baseline covariates available before treatment.  Choose an RMST restriction time $t^\star$ before fitting the model.  For a patient with baseline profile $x$, define
\begin{equation}
 \Delta(x)=\mathrm{RMST}_1(t^\star\mid x)-\mathrm{RMST}_0(t^\star\mid x),
 \label{eq:delta}
\end{equation}
where the subscripts 1 and 0 refer to the two randomized treatment groups.  Thus $\Delta(x)$ is the conditional RMST difference through $t^\star$.  A positive value means that the patient profile has greater restricted survival time under treatment 1.

We estimate a score $r(x)$ whose ordering agrees with that of $\Delta(x)$. Thus, $\Delta(x_1)>\Delta(x_2)$ should imply $r(x_1)>r(x_2)$. The numerical value of $r(x)$ is not interpreted as an RMST difference.

\subsection{Estimation}

Event and censoring models are fitted with sample splitting. For each patient, these models provide a fitted conditional RMST difference and a censoring-adjusted RMST pseudo outcome, denoted by $Y_i^*$. At the population values of the event and censoring functions,
\begin{equation}
 \E(Y_i^*\mid X_i=x)=\Delta(x).
 \label{eq:pseudomean}
\end{equation}
Thus, $Y_i^*$ supplies an adjusted patient-level outcome whose conditional mean is the target treatment effect. Its expression is given in Appendix~\ref{app:pseudo}.

A regression approach fits the conditional mean of $Y_i^*$ and ranks the fitted values. Here, the fitted RMST differences first determine a smooth comparison for each pair $(i,j)$. The residual terms $Y_i^*-\widehat\Delta_i$ and $Y_j^*-\widehat\Delta_j$ then correct this comparison. The resulting pairwise targets are used in a logistic loss for $r(x)$. Appendix~\ref{app:pairwise} gives the formulas. The score may be linear, penalized, or fitted with a more flexible function class, depending on the available sample size.

Algorithm~\ref{alg:ranking} summarizes the analysis.

\begin{algorithm}[H]
\caption{Estimation of an RMST treatment benefit ranking score}
\label{alg:ranking}
\begin{algorithmic}[1]
\Require Randomized trial data, baseline covariates, a prespecified RMST restriction time, and a model form for the final score.
\State Split the data into folds.
\State Using the other folds, fit the event and censoring models and calculate, for each held out patient, a fitted RMST difference and the RMST pseudo outcome $Y_i^*$.
\State Repeat Step 2 until every patient has quantities obtained without using that patient's own outcome to fit the survival models.
\State Form patient pairs.  For each pair, construct a soft comparison of which patient has the larger RMST benefit and correct that comparison using the two RMST pseudo outcomes.
\State Fit the score $r(X)$ with a pairwise logistic loss so that patients favored by the corrected comparisons receive larger scores.
\State Apply $r(X)$ to new patients.  Assess the score by comparing treatment effects across score groups or among patients with the largest scores.
\end{algorithmic}
\end{algorithm}

RMST regression and direct ranking use the same survival models and censoring-adjusted pseudo outcomes. Their difference lies in the final loss: squared error for RMST regression and pairwise logistic loss for direct ranking.

The pairs have a different role from those in rank sum tests, generalized pairwise comparisons, and the win ratio. In those procedures, observed outcomes from patients in different treatment groups contribute to a marginal treatment comparison. Here, a pair consists of two baseline profiles, and the comparison concerns their conditional RMST benefits. Treatment assignment and censoring are handled in the construction of $Y_i^*$ before the pairwise loss is formed.

\subsection{Statistical properties}

Appendix~\ref{app:pairwise} defines the pairwise target and population loss. The following results apply to the ideal population criterion with a fixed $\kappa>0$ and without finite-sample truncation of the corrected pairwise target.

\begin{proposition}[Population target of the ranking score]
\label{prop:ordering}
Assume consistency, randomization, conditionally independent censoring, and positivity through $t^\star$. Suppose the event and censoring distributions are evaluated at their population values and the score class contains the population minimizer of the pairwise loss. Then
\begin{equation}
 r_0(x)=\frac{\Delta(x)}{\kappa}+c,
 \label{eq:popscore}
\end{equation}
where $c$ is a constant. Consequently, for almost every pair of baseline profiles,
\begin{equation}
 \Delta(x)>\Delta(x')
 \quad\Longleftrightarrow\quad
 r_0(x)>r_0(x').
 \label{eq:ordering}
\end{equation}
\end{proposition}

Proposition~\ref{prop:ordering} shows that the population score follows the ordering of conditional treatment benefit. Its scale depends on $\kappa$, and the additive constant is unidentified, so the fitted score is used for ranking. The proof is given in Appendix~\ref{app:proof1}.

The second result addresses estimation of the event and censoring distributions.  For its formal statement, let $\eta$ denote these nuisance functions and let $\mathcal R(r,\eta)$ denote the population pairwise risk.  Write $\eta_0$ for the population values of the nuisance functions.

\begin{theorem}[Neyman orthogonality of the pairwise estimating equation]
\label{thm:orth}
Under the conditions of Proposition~\ref{prop:ordering} and the regularity conditions for the orthogonal RMST construction of \citet{frauen2025}, let $r_0$ be a population minimizer and treat $\kappa$ as fixed. For every admissible perturbation $h_\eta$ of the event and censoring functions and every square-integrable perturbation $h_r$ of the ranking score,
\begin{equation}
 D_\eta D_r\mathcal R(r_0,\eta_0)[h_r,h_\eta]=0.
 \label{eq:mainorth}
\end{equation}
\end{theorem}

Here $D_r$ and $D_\eta$ denote directional derivatives with respect to the score and the nuisance functions. Equation~\eqref{eq:mainorth} states that local errors in the event and censoring functions have no first-order effect on the population estimating equation for $r$. Appendix~\ref{app:proof2} gives the derivation.

Both RMST regression and direct ranking recover the correct population ordering under sufficiently rich models. Their finite-sample performance can differ because their losses emphasize different errors. The simulations examine this difference when performance is evaluated by ranking and treatment-benefit enrichment.

\subsection{Evaluation of a fitted ranking}

We evaluate whether patients with higher scores have larger treatment benefits using the targeting operator characteristic \citep{yadlowsky2025}. For a score $r(X)$ and a fraction $q\in(0,1]$, let $c_q$ be the cutoff that selects the highest-scoring fraction $q$ of the population. The TOC at $q$ is
\begin{equation}
 \begin{aligned}
 \mathrm{TOC}_r(q)
 &=\E\{\Delta(X)\mid r(X)\ge c_q\}\\
 &\quad-\E\{\Delta(X)\}.
 \end{aligned}
 \label{eq:toc}
\end{equation}
It measures the increase in mean treatment benefit among the highest-scoring fraction $q$ relative to the full population. We refer to $\mathrm{TOC}_r(q)$ at a specified value of $q$ as top-$q$ enrichment. The area under the TOC curve is
\begin{equation}
 \mathrm{AUTOC}(r)=\int_0^1\mathrm{TOC}_r(q)\,dq.
 \label{eq:autoc}
\end{equation}
A larger AUTOC indicates that patients with greater treatment benefit tend to receive higher scores. In the simulations, these quantities are calculated from the known $\Delta(X)$. In the data analysis, they are estimated with out-of-fold scores and RMST pseudo outcomes calculated on the evaluation observations.
Specifically, let $\mathcal I_q$ index the $n_q$ patients in the highest-scoring fraction $q$, let $\overline Y_q^*$ be the mean of $Y_i^*$ among these patients, and let $\overline Y^*$ be the overall mean. The empirical TOC is
\begin{equation}
 \widehat{\mathrm{TOC}}_r(q)=\overline Y_q^*-\overline Y^*,
 \label{eq:tochat}
\end{equation}
and the empirical AUTOC is obtained by integrating this curve over the evaluated fractions $q$.

\section{Simulation Study}

The simulations used a monotone nonlinear relationship between a latent response score and the conditional RMST difference. The settings also included covariate-dependent censoring and a working event model that omitted a treatment-effect interaction. In each replicate of the main setting, the three methods were fitted in a training sample of 300 patients randomized 1:1. The data included eight baseline covariates, and approximately 34\% of event times were censored. Treatment benefit depended on three covariates and one interaction. The working event model omitted the interaction, while the final RMST regression and ranking models both included it and used the same cross-fitted pseudo outcome. The data-generating mechanism is given in Appendix~\ref{app:sim}.

Three methods were compared. The RMST plug-in method ranked the conditional RMST differences estimated by the working survival model. RMST regression applied squared-error regression to the cross-fitted RMST pseudo outcome and ranked the fitted values. Direct ranking used the same pseudo outcome and final covariates but fitted the pairwise criterion described in Section 2. In each replicate, performance was evaluated in a separately generated test sample of 1200 patients; these patients were not used for model fitting or tuning. Spearman and Kendall correlations measured agreement with the true ordering of \(\Delta(X)\). The top-20\% benefit was the mean true treatment benefit among patients receiving the highest 20\% of scores. Regret was the mean benefit in the true top 20\% minus the top-20\% benefit selected by a fitted score, so smaller values are better. AUTOC was calculated as in \eqref{eq:autoc}. The main setting was repeated 500 times. Three additional settings examined heavier censoring, stronger nonlinearity, and a smaller sample with heavy censoring; each was repeated 150 times.

\begin{table}[t]
\centering
\caption{Simulation results. Entries are Monte Carlo means. Higher values are better except for regret. The main setting used 500 repetitions; the other settings used 150 repetitions. Bold type marks the best result within each setting.}
\label{tab:sim}
\small
\setlength{\tabcolsep}{4.5pt}
\begin{tabular}{llrrrrr}
\toprule
Setting & Method & Spearman & Kendall & Top 20\% & Regret & AUTOC\\
\midrule
Main & RMST plug-in & 0.467 & 0.330 & 0.468 & 0.585 & 0.291\\
 & RMST regression & 0.509 & 0.368 & 0.531 & 0.521 & 0.316\\
 & Direct ranking & \textbf{0.541} & \textbf{0.392} & \textbf{0.574} & \textbf{0.479} & \textbf{0.337}\\[2pt]
Heavy censoring & RMST plug-in & 0.401 & 0.283 & 0.388 & 0.671 & 0.251\\
 & RMST regression & 0.455 & 0.325 & 0.465 & 0.594 & 0.283\\
 & Direct ranking & \textbf{0.494} & \textbf{0.355} & \textbf{0.519} & \textbf{0.541} & \textbf{0.309}\\[2pt]
Stronger nonlinearity & RMST plug-in & 0.342 & 0.239 & 0.341 & 0.745 & 0.221\\
 & RMST regression & 0.463 & 0.335 & 0.495 & 0.590 & 0.295\\
 & Direct ranking & \textbf{0.494} & \textbf{0.356} & \textbf{0.537} & \textbf{0.548} & \textbf{0.316}\\[2pt]
Small sample, heavy censoring & RMST plug-in & 0.306 & 0.214 & 0.280 & 0.789 & 0.194\\
 & RMST regression & 0.386 & 0.275 & 0.384 & 0.685 & 0.242\\
 & Direct ranking & \textbf{0.419} & \textbf{0.298} & \textbf{0.423} & \textbf{0.646} & \textbf{0.263}\\
\bottomrule
\end{tabular}
\end{table}

Direct ranking had the highest mean rank correlations, top-20\% benefit, and AUTOC, and the lowest regret in all four settings. In the main setting, its mean Spearman correlation was 0.541, compared with 0.509 for RMST regression and 0.467 for the RMST plug-in method. Relative to RMST regression, the paired mean difference was 0.032 (Monte Carlo standard error 0.005). The corresponding difference in AUTOC was 0.021 (0.003), and top-20\% regret was lower by 0.042 (0.007). Under heavier censoring, the Spearman correlations were 0.401, 0.455, and 0.494 for the RMST plug-in method, RMST regression, and direct ranking, respectively. The corresponding mean benefits in the selected top 20\% were 0.388, 0.465, and 0.519. All three methods performed less well in the smaller sample with heavy censoring, with direct ranking retaining the highest mean rank correlations and AUTOC.

\begin{figure}[t]
\centering
\includegraphics[width=0.83\textwidth]{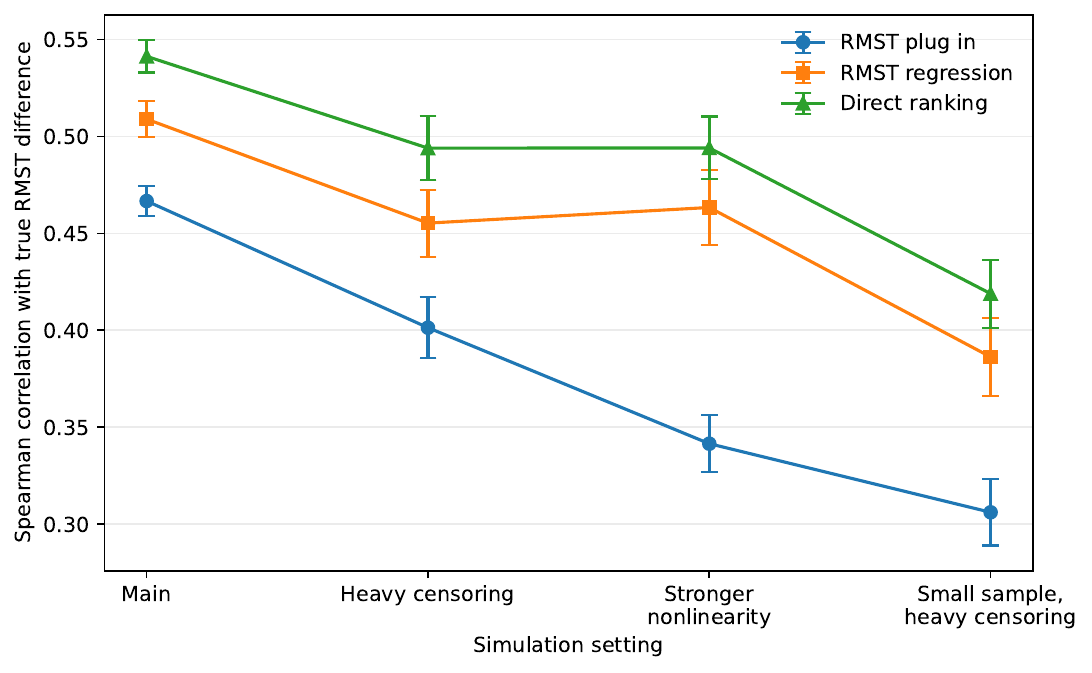}
\caption{Spearman correlation between each ranking score and the true conditional RMST difference. Error bars are Monte Carlo standard errors.}
\label{fig:sim}
\end{figure}

\section{Application to a Randomized Breast Cancer Trial}

We applied the three methods to data from the German Breast Cancer Study Group 2.  The original study was a multicenter randomized clinical trial evaluating hormonal treatment with tamoxifen and the duration of adjuvant chemotherapy in women with node-positive breast cancer \citep{schumacher1994}.  The publicly available data contain 686 women with complete baseline information and recurrence-free survival follow-up \citep{sauerbrei1999}.  Hormonal therapy was recorded for 242 women and no hormonal therapy for 444.  The baseline variables were age, menopausal status, tumor size, tumor grade, number of positive lymph nodes, progesterone receptor level, and estrogen receptor level.

The analysis compared hormonal therapy with no hormonal therapy using recurrence-free survival through three years. The restriction time was chosen before fitting the ranking models. At three years, 197 women without hormonal therapy and 136 women receiving hormonal therapy remained at risk. We used out-of-fold prediction for every reported score and tuned the direct ranking method within the corresponding training data. The event model, censoring model, RMST regression, and direct ranking model used the same baseline variables. Appendix~\ref{app:gbsg} gives the implementation details.

The estimated overall three-year RMST difference was 62.5 days in favor of hormonal therapy. Direct ranking had the largest area under the TOC curve (AUTOC) point estimate, 51.6 days, followed by RMST regression at 47.6 days and the RMST plug-in score at 38.0 days (Table~\ref{tab:real}). The paired AUTOC difference between direct ranking and RMST regression was 4.0 days, with a descriptive bootstrap interval of $-10.9$ to 19.6 days.

At $q=0.30$, the estimated TOC, or top-30\% enrichment, was 63.4 days for direct ranking, 50.0 days for the RMST plug-in score, and 33.7 days for RMST regression. All three methods were evaluated on the same patients using out-of-fold scores and the same evaluation pseudo outcomes.

\begin{table}[t]
\centering
\caption{Evaluation in the GBSG2 breast cancer data using recurrence-free survival through three years. Values are in days. Intervals are descriptive bootstrap intervals conditional on the out-of-fold scores. Bold type marks the largest AUTOC and top-30\% enrichment point estimates.}
\label{tab:real}
\small
\begin{tabular}{lrrr}
\toprule
Method & AUTOC (95\% interval) & Top 30\% enrichment & Overall effect\\
\midrule
RMST plug-in & 38.0 ($-14.1$, 87.2) & 50.0 & 62.5\\
RMST regression & 47.6 ($-3.3$, 95.6) & 33.7 & 62.5\\
Direct ranking & \textbf{51.6} (2.2, 100.4) & \textbf{63.4} & 62.5\\
\bottomrule
\end{tabular}
\end{table}

\section{Discussion}

We developed a pairwise procedure for ranking patients by conditional RMST benefit in randomized trials with right-censored outcomes. The construction combines an orthogonal RMST pseudo outcome with a smooth treatment-effect ranking loss. At the population level, the resulting score has the same ordering as the conditional RMST difference. The estimating equation is also locally insensitive to errors in the fitted event and censoring distributions.
In the simulations, direct ranking improved rank correlation and treatment-benefit enrichment relative to plug-in estimation and regression of the same pseudo outcome. The breast cancer analysis gave the largest AUTOC and top-30\% enrichment point estimates for direct ranking.

\appendix

\section{Formal Definition of the Procedure}

\subsection{RMST pseudo outcome}
\label{app:pseudo}

This section gives the RMST pseudo outcome used in computation and theory. Represent time on a grid $k=0,\ldots,K$, with the final grid point corresponding to $t^\star$. Let
\[
\begin{aligned}
 S_k(x,a)&=\Prb(T>k\mid X=x,A=a),\\
 G_k(x,a)&=\Prb(C>k\mid X=x,A=a),
\end{aligned}
\]
and let $\lambda_k^S(x,a)$ denote the event hazard. Randomization implies that $\pi=\Prb(A=1\mid X)$ is known. We assume consistency, randomization, conditionally independent censoring given $(X,A)$, and positivity through $K$.

Following the unweighted RMST construction of \citet{frauen2025}, define the event martingale increment
\[
\begin{aligned}
 M_u(Z)={}&\ind(\widetilde T=u,\Delta_T=1)\\
 &-\ind(\widetilde T\ge u)\lambda_u^S(X,A).
\end{aligned}
\]
Then
\begin{equation}
 \xi_{S,k}(Z;\eta_k)
 =\sum_{u=0}^{k}\frac{M_u(Z)}
 {S_u(X,A)G_{u-1}(X,A)},
 \label{eq:xi}
\end{equation}
with $G_{-1}\equiv1$.  The RMST pseudo outcome is
\begin{equation}
 \begin{aligned}
 Y^*(Z;\eta)=\sum_{k=0}^{K}\Bigg[&S_k(X,1)-S_k(X,0)\\
 &-\frac{A-\pi}{\pi(1-\pi)}S_k(X,A)
   \xi_{S,k}(Z;\eta_k)\Bigg].
 \end{aligned}
 \label{eq:ystar}
\end{equation}
At the population values, $\E\{Y^*(Z;\eta_0)\mid X=x\}=\Delta(x)$.  Equation~\eqref{eq:ystar} is the $f\equiv1$ RMST specialization of the orthogonal survival construction in Appendix F.2 of \citet{frauen2025}.

Up to terms that do not depend on a regression function $m$, the corresponding squared error loss can be written as
\[
 L_{\rm RMST}(m,\eta)=\E\left[m(X)^2-2m(X)Y^*(Z;\eta)\right].
\]
At the population minimizer, orthogonality of this loss gives
\begin{equation}
 \E\left[a(X)D_\eta Y^*(Z;\eta_0)[\dot\eta]\right]=0
 \label{eq:ystarorth}
\end{equation}
for square integrable $a(X)$ and admissible perturbations $\dot\eta$.

\subsection{Pairwise target and loss}
\label{app:pairwise}

Let $\eta$ collect the event and censoring functions, and let $\Delta_\eta(x)$ be the RMST contrast calculated from the event distribution in $\eta$. For two independent observations $Z_i$ and $Z_j$, write
\[
 \delta_{ij}(\eta)=\Delta_\eta(X_i)-\Delta_\eta(X_j).
\]
Define
\begin{equation}
 \begin{aligned}
 q_{ij}(\eta)
 &=\expit\{\delta_{ij}(\eta)/\kappa\},\\
 \omega_{ij}(\eta)
 &=\frac{q_{ij}(\eta)\{1-q_{ij}(\eta)\}}{\kappa},
 \end{aligned}
 \label{eq:q}
\end{equation}
where $\kappa>0$ controls the scale of the smooth comparison. Also let
\[
 e_i(\eta)=Y^*(Z_i;\eta)-\Delta_\eta(X_i).
\]
The corrected pairwise target is
\begin{equation}
 \widetilde q_{ij}(\eta)
 =q_{ij}(\eta)+\omega_{ij}(\eta)
 \{e_i(\eta)-e_j(\eta)\}.
 \label{eq:pseudolabel}
\end{equation}
For a ranking score $r$, write
\[
 p_r(X_i,X_j)=\expit\{r(X_i)-r(X_j)\}.
\]
The population pairwise loss is
\begin{equation}
 \begin{aligned}
 \mathcal R(r,\eta)=\E\Big[&-\widetilde q_{ij}(\eta)\log p_r\\
 &-\{1-\widetilde q_{ij}(\eta)\}\log(1-p_r)\Big].
 \end{aligned}
 \label{eq:risk}
\end{equation}
In the empirical loss, $\eta$ is replaced by cross-fitted estimates. The resulting corrected targets are truncated to a small interval inside $(0,1)$ for numerical stability. The population results below concern the untruncated target with fixed $\kappa$. In the data example, $\kappa$ was selected within the training data.

\subsection{Proof of Proposition~\ref{prop:ordering}}
\label{app:proof1}

At the population values,
\[
 \E\{Y^*(Z;\eta_0)-\Delta(X)\mid X\}=0.
\]
For independent $i$ and $j$, conditioning on $(X_i,X_j)=(x,x')$ gives
\[
 \E\{\widetilde q_{ij}(\eta_0)\mid x,x'\}
 =q_{ij}(\eta_0)
 =\expit\left\{\frac{\Delta(x)-\Delta(x')}{\kappa}\right\}.
\]
Conditional cross entropy is affine in the soft target and strictly convex in the pairwise probability.  Hence it is minimized at $p_r(x,x')=q_{ij}$.  Because the logistic map is one to one,
\[
 r(x)-r(x')=\frac{\Delta(x)-\Delta(x')}{\kappa}
\]
for almost every pair.  Under the usual connectedness condition on the support of $X$, this implies
\[
 r(x)=\Delta(x)/\kappa+c,
\]
for a constant $c$, and therefore gives the ordering stated in Proposition~\ref{prop:ordering}.

\subsection{Proof of Theorem~\ref{thm:orth}}
\label{app:proof2}

The formal statement corresponding to Theorem~\ref{thm:orth} is
\begin{equation}
 D_\eta D_r\mathcal R(r^0,\eta_0)[\dot r,\dot\eta]=0,
 \label{eq:orthformal}
\end{equation}
where $r^0$ is a population minimizer and $\dot r,\dot\eta$ are admissible perturbation directions.

Let $d_r(X_i,X_j)=\dot r(X_i)-\dot r(X_j)$.  Differentiating the pairwise cross entropy with respect to the score gives
\[
 D_r\mathcal R(r,\eta)[\dot r]
 =\E\left[\{p_r(X_i,X_j)-\widetilde q_{ij}(\eta)\}d_r(X_i,X_j)\right].
\]
Because $p_r$ does not depend on $\eta$,
\begin{equation}
 \begin{aligned}
 &D_\eta D_r\mathcal R(r^0,\eta_0)[\dot r,\dot\eta]\\
 &\quad=-\E\left[D_\eta\widetilde q_{ij}(\eta_0)[\dot\eta]
 d_r(X_i,X_j)\right].
 \end{aligned}
 \label{eq:mixed}
\end{equation}
Since
\[
 \frac{\partial q_{ij}(\eta_0)}{\partial\delta_{ij}(\eta_0)}
 =\omega_{ij}(\eta_0),
\]
the derivative of the plug-in component of $\widetilde q_{ij}$ is
\[
 \begin{aligned}
 \omega_{ij}(\eta_0)\{&D_\eta\Delta_\eta(X_i)
 -D_\eta\Delta_\eta(X_j)\}\big|_{\eta=\eta_0}.
 \end{aligned}
\]
At $\eta_0$, $\E\{Y^*(Z_i;\eta_0)-\Delta_{\eta_0}(X_i)\mid X_i\}=0$. Terms arising from differentiation of $\omega_{ij}(\eta)$ therefore vanish after conditioning on $(X_i,X_j)$. The remaining derivative of the correction term is
\[
 \omega_{ij}(\eta_0)
 \{D_\eta e_i(\eta)-D_\eta e_j(\eta)\}
 \big|_{\eta=\eta_0}.
\]
The derivatives of $\Delta_\eta(X_i)$ and $\Delta_\eta(X_j)$ cancel those from the plug-in component. By symmetry and iterated expectation, the remaining contribution to \eqref{eq:mixed} is a linear combination of terms of the form
\[
 \E\left[a(X)D_\eta Y^*(Z;\eta_0)[\dot\eta]\right],
\]
which are zero by \eqref{eq:ystarorth}.  This proves \eqref{eq:orthformal}.

\section{Simulation Details}
\label{app:sim}

Eight independent baseline covariates were generated from a standard normal distribution.  A latent response score was
\[
 s(X)=0.95X_1-0.70X_2+0.75X_3+\gamma X_1X_3,
\]
and the conditional RMST difference was
\begin{equation}
 \Delta(X)=0.95\tanh\{s(X)\}+0.056s(X).
 \label{eq:simdelta}
\end{equation}
Because \eqref{eq:simdelta} is strictly increasing in $s(X)$, the response score determines the correct ordering of treatment benefit while its numerical magnitude remains nonlinear.  The control RMST was
\[
 \mu_0(X)=3.05+0.28\tanh(0.65X_4-0.45X_5+0.25X_6),
\]
and $\mu_1(X)=\mu_0(X)+\Delta(X)$.  Treatment was randomized 1:1.  Geometric event hazards were chosen so that these quantities were the exact RMST values over six discrete time intervals.

Censoring depended on $X_7$ and $X_8$ through
\[
 \lambda_C(X)=\expit\{\alpha_C+\beta_C(0.9X_7+0.8X_8)\}.
\]
Event and censoring hazards were estimated with pooled logistic working models containing the eight main covariate terms. The $X_1X_3$ interaction was omitted from the event model. RMST regression and direct ranking used the same final covariates $(X_1,\ldots,X_8,X_1X_3)$. Three-fold sample splitting was used to construct the pseudo outcomes in the training sample. Each fitted score was evaluated in an independently generated test sample of 1200 observations.

The main setting used a training sample size of $n=300$, $\gamma=1.10$, and approximately 34\% censoring and was repeated 500 times. Three additional settings considered heavier censoring, stronger nonlinear effect modification, and a smaller sample with heavy censoring. Each additional setting used 150 repetitions.

In the main setting, the Monte Carlo standard errors of the mean Spearman correlations were 0.0076 for RMST plug-in estimation, 0.0091 for RMST regression, and 0.0083 for direct ranking. For the paired comparison of direct ranking with RMST regression, the mean difference in Spearman correlation was 0.0323 with Monte Carlo standard error 0.0051. The mean difference in AUTOC was 0.0214 with Monte Carlo standard error 0.0032, and top-20\% regret was lower by 0.0421 with Monte Carlo standard error 0.0066.

Implementation used 7000 ordered pairs per training sample.  Corrected pairwise targets were truncated to $[0.001,0.999]$, and an $L_2$ penalty was applied to the pairwise logistic coefficients.  The estimated censoring survival in the martingale denominator was bounded below at 0.08 for numerical stability.

\section{Details of the Breast Cancer Trial Analysis}
\label{app:gbsg}

The GBSG2 data used in the application contain 686 women with recurrence-free survival follow-up and no missing values in the baseline variables used for modeling. Hormonal therapy was coded as the randomized treatment of interest. The baseline model included age, menopausal status, tumor size, tumor grade, number of positive lymph nodes, progesterone receptor, and estrogen receptor. Continuous variables were transformed on clinically reasonable log scales where appropriate, and a small number of prespecified quadratic and interaction terms were included to allow treatment benefit to vary nonlinearly with receptor status and disease burden.

The restriction time was three years. Event and censoring hazards were estimated on a discrete time grid using pooled logistic models. We used repeated outer sample splitting so that every reported patient score was obtained from models that did not use that patient's outcome. Within each outer training sample, the RMST pseudo outcomes used to fit RMST regression and direct ranking were themselves obtained by cross-fitting. The smoothness parameter for direct ranking was selected within the outer training data using an inner validation split and an AUTOC criterion. Model fitting, tuning, and evaluation used separate data splits.

At three years, 197 women without hormonal therapy and 136 women receiving hormonal therapy remained at risk. The evaluation pseudo outcome gave an overall RMST difference of approximately 63 days. AUTOC values were 38.0, 47.6, and 51.6 days for RMST plug-in estimation, RMST regression, and direct ranking, respectively. In a paired descriptive bootstrap, the AUTOC difference between direct ranking and RMST regression was 4.0 days, with a 95\% interval from $-10.9$ to 19.6 days. The largest separation in the point estimates occurred near the upper part of the ranking: top-30\% enrichment was 63.4 days for direct ranking and 33.7 days for RMST regression. These uncertainty calculations condition on the fitted out-of-fold scores and therefore do not represent full refitting bootstrap confidence intervals.

\bibliographystyle{plainnat}
\bibliography{ref}

\end{document}